\documentclass{appolb}

\usepackage{calc}
\usepackage{graphicx}
\usepackage{dcolumn}
\usepackage{bm}
\usepackage{slashed}
\usepackage{amsmath,graphicx}
\usepackage[colorlinks=true,linktocpage=true,linkcolor=blue,citecolor=blue]{hyperref}
\usepackage{float}
\usepackage{nicefrac}
\usepackage[normalem]{ulem}
\usepackage{amsmath}
\usepackage{subfigure}
\usepackage{float} 
\usepackage{bbold}
\usepackage{cite} 

\def\Cv{{\boldsymbol C}}

\def\nU{n_{(0)}}
\def\eU{\varepsilon_{(0)}}
\def\PU{P_{(0)}}
\def\sU{s_{(0)}}

\newcommand{\sh}[1]{\sinh#1}
\newcommand{\ch}[1]{\cosh#1}

\def\HP{\hphantom{\alpha}} 

\newcommand{\beq}{\begin{eqnarray}}
\newcommand{\eeq}{\end{eqnarray}}

\newcommand{\bea}{\begin{eqnarray}}
\newcommand{\eea}{\end{eqnarray}}

\newcommand{\bel}[1]{\begin{eqnarray}\label{#1}}
\newcommand{\eel}{\end{eqnarray}}

\def\a{\alpha}
\def\b{\beta}
\def\g{\gamma}
\def\d{\delta} 

\def\s{\sigma}

\newcommand{\rf}[1]{Eq.~(\ref{#1})}
\newcommand{\rfm}[1]{Eqs.~(\ref{#1})}

\newcommand{\rfn}[1]{(\ref{#1})}

\newcommand{\nn}{\nonumber}

\newcommand{\p}{\partial}

\newcommand{\f}[2]{\frac{#1}{#2}}
\newcommand{\onehalf}{{\nicefrac{1}{2}}}

\renewcommand\sout{\bgroup \color{blue} \ULdepth=-.5ex \ULset}

\def\n0{n_{(0)}}
\def\e0{\varepsilon_{(0)}}
\def\P0{P_{(0)}}
\def\s0{s_{(0)}}

\def\LR{\left(} 
\def\RR{\right)}
\def\LS{\left[} 
\def\RS{\right]}

\def\HP{\hphantom{\alpha}} 

\def\fplusrsxp{f^+_{rs}(x,p)}

\def\fminusrsxp{f^-_{rs}(x,p)}

\def\bmu{\beta_\mu}

\def\omnL{\omega_{\mu\nu}}
\def\omnU{\omega^{\mu\nu}}

\def\pmu{p^\mu}

\def\pv{{\boldsymbol p}}

\def\SmunuU{{\Sigma}^{\mu\nu}}

\def\S0iU{{\Sigma}^{0i}} 
 
\def\SmnU{{\Sigma}^{\mu\nu}}

\def\CHI{\chi}

\def\ubarrp{{\bar u}_r(p)}

\def\usp{u_s(p)}
\def\urp{u_r(p)}

\def\vbarrp{{\bar v}_r(p)}
\def\vbarsp{{\bar v}_s(p)}
\def\vsp{v_s(p)}
\def\vrp{v_r(p)}

\def\g5{\gamma_5}

\def\Weqpmxk{{\cal W}^{\pm}_{\rm eq}(x,k)}

\def\Feqpmxk{{\cal F}^{\pm}_{\rm eq}(x,k)}

\def\Peqpmxk{{\cal P}^{\pm}_{\rm eq}(x,k)}

\begin{document} 
\title{
Hydrodynamics formalism with Spin dynamics
\thanks{Presented at the on-line meeting Criticality in QCD and the Hadron Resonance Gas, Wrocław, Poland, July 29–31, 2020} 
}
\author{Rajeev Singh
		 \address{ 
		 	Institute of Nuclear Physics Polish Academy of Sciences,\\ PL-31342 Krak\'ow, Poland}
}
\maketitle
\begin{abstract} 
We review the key steps of the relativistic fluid dynamics formalism with spin degrees of freedom initiated recently. We obtain equations of motion of the expansion of the system from the underlying definitions of quantum kinetic theory for the equilibrium phase space distribution functions. We investigate the dynamics of spin polarization of the system in the Bjorken hydrodynamical background.
\end{abstract}
%
%
\section{Introduction}
%
%
Spin polarization experimental measurements of $\Lambda$ hyperons recently taken by the STAR Collaboration \cite{STAR:2017ckg,Adam:2018ivw,Niida:2018hfw,Adam:2019srw} have created a huge interest in the spin polarization studies and in studies correlating between the vorticity and particle spin polarization in relativistic heavy-ion collisions\cite{Shi:2020htn,Weickgenannt:2020aaf,Becattini:2009wh,Becattini:2013fla,Montenegro:2017rbu,Montenegro:2017lvf,Becattini:2018duy,Boldizsar:2018akg,Prokhorov:2018bql,Yang:2018lew,Florkowski:2019voj,Weickgenannt:2019dks,Hattori:2019lfp,Ambrus:2019ayb,Sheng:2019kmk,Prokhorov:2019cik,Ivanov:2019wzg,Hattori:2019ahi,Xie:2019jun,Liu:2019krs,Wu:2019eyi,Becattini:2019ntv,Zhang:2019xya,Li:2019qkf,Florkowski:2019gio,Prokhorov:2019yft,Fukushima:2020qta,Liu:2020ymh,Bhadury:2020puc,Tabatabaee:2020efb,Liu:2020flb,Yang:2020hri,Deng:2020ygd,Taya:2020sej,Gao:2020lxh,Bhadury:2020cop,Montenegro:2020paq}; for reviews see\cite{Becattini:2020ngo,Tinti:2020gyh,Speranza:2020ilk,Gao:2020vbh,Becattini:2020sww}. Thermal-based models~\cite{Becattini:2016gvu,Karpenko:2016jyx,Li:2017slc,Xie:2017upb} which precisely explain the global polarization of particles, does able to explain differential results correctly\cite{Adam:2019srw}, these models assume the condition that particle spin polarization emitted at freeze-out hypersurface is defined by the thermodynamical quantity which is named as thermal vorticity~\cite{Becattini:2007sr,Becattini:2013fla}, not considering the fact that it may evolve independently during the expansion of the fluid. In this article, we follow scheme proposed in Refs.~\cite{Florkowski:2017ruc,Florkowski:2017dyn,Florkowski:2018myy,Florkowski:2018fap,Florkowski:2018ahw,Florkowski:2019qdp}, and analyze such possibility of spin polarization evolution using relativistic hydrodynamics framework with spin.
\section{Distribution functions in equilibrium}
If we know phase space distribution function for the system's equilibrium state, then it is possible to derive relativistic hydrodynamics from the underlying kinetic theory definitions\cite{Florkowski:2017olj}. Following ideas developed by Becattini \textit{et al}.\cite{Becattini:2013fla}, we take into consideration the following distribution functions for the relativistic systems of spin $\onehalf$ massive particles (and antiparticles) in the local equilibrium state.
\bel{fplusrsxp}
\fplusrsxp =  \ubarrp X^+ \usp, \qquad
\fminusrsxp = - \vbarsp X^- \vrp,
\eel
where $x$ and $p$ is the space-time position coordinate and the four-vector momentum, respectively, with $\urp$ and $\vrp$ being the Dirac bispinors ($r,s = 1,2$). Dirac bispinors follow the normalization conditions as $\ubarrp \usp=\,\delta_{rs}$ and $\vbarrp \vsp=-\,\delta_{rs}$, here $\delta_{rs}$ is the Dirac delta function and, $X^{\pm}$ have the following form in terms of relativistic Boltzmann distributions
\bel{XpmM}
X^{\pm} =  \exp\left[\pm \xi(x) - \bmu(x) \pmu  \pm \f{1}{2} \omnL(x)  \SmunuU \right], \nn
\eel
where $\beta^\mu \equiv U^\mu/T$ and $\xi \equiv \mu/T$, here $T$, $\mu$ and $U^\mu$ is the temperature, baryon chemical potential and four-vector velocity,  respectively, and $\omnL$ is the second rank asymmetric tensor known as spin polarization tensor with $\SmunuU  \equiv  \f{i}{4} [\gamma^\mu,\gamma^\nu]$ being the spin operator.  

With the help of the expressions from Ref.~\cite{DeGroot:1980dk} and Eqs. \rfn{fplusrsxp} we can obtain the Wigner functions in equilibrium as follows
\bea
{\cal W}^\pm_{\rm eq}(x,k) &=& \frac{e^{\pm \xi}}{4 m}  \int dP
\,e^{-\beta \cdot p }\,\, \delta^{(4)}(k \mp p) \label{eq:Weqpxk2} \\
&& \times  \left[2m (m \pm \slashed{p}) \cosh(\zeta) \pm \f{\sinh(\zeta)}{2\zeta}  \, \omnL \,(\slashed{p} \pm m) \SmnU (\slashed{p} \pm m) \right],\nn
\eea
where $k$ being the off mass-shell particle four-momentum, 
$dP = d^3p/((2 \pi )^3 E_p)$ is the invariant measure with $E_p = \sqrt{m^2 + \pv^2}$ denoting the on-shell energy of the particle, and $\zeta =  \f{1}{2 \sqrt{2}} \sqrt{ \omnL \omnU}$ in terms of spin polarization tensor.\\
Wigner function can also be expanded using Clifford-algebra expansion \rfn{eq:Weqpxk2}
\bea
\Weqpmxk &=& \f{1}{4} \left[ \Feqpmxk + i \gamma_5 \Peqpmxk + \gamma^\mu {\cal V}^\pm_{{\rm eq}, \mu}(x,k) \right. \nn \\
&& \left.  \hspace{1cm} + \gamma_5 \gamma^\mu {\cal A}^\pm_{{\rm eq}, \mu}(x,k)
+ \SmnU {\cal S}^\pm_{{\rm eq}, \mu \nu}(x,k) \right], \nn
\label{eq:wig_expansion}
\eea
where $\mathcal{X} \in\left\{\mathcal{F}, \mathcal{P}, \mathcal{V}_{\mu}, \mathcal{A}_{\mu}, \mathcal{S}_{\nu \mu}\right\}$ are the coefficient functions of the Wigner function, which can obtained from the trace of $\Weqpmxk$ multiplying first by: $\left\{\mathbf{1},-i \gamma_{5}, \gamma_{\mu}, \gamma_{\mu} \gamma_{5}, 2 \Sigma_{\mu \nu}\right\}$.
%
%
\section{Kinetic and hydrodynamical equations}
The kinetic equation to be followed by Wigner function is
\bel{eq:eqforWC}
\left(\gamma_\mu K^\mu - m \right) {\cal W}(x,k) = C[{\cal W}(x,k)],
\eel
with $K^\mu = k^\mu + \frac{i \hbar}{2} \,\p^\mu$. 
For global equilibrium state, the Wigner function follows exactly \rf{eq:eqforWC} with the collision term $C[{\cal W}(x,k)] = 0$. The widely used method of treating \rf{eq:eqforWC} is the semi-classical expansion method of the coefficient functions of the Wigner function 
\bel{eq:semi}
\mathcal{X}=\mathcal{X}^{(0)}+\hbar \mathcal{X}^{(1)}+\hbar^{2} \mathcal{X}^{(2)}+\cdots. \nn
\eel
Up to the first order (i.e. next-to-leading order) in $\hbar$ the treatment of \rf{eq:eqforWC} gives the following kinetic equations to be followed by the two independent coefficients which are: $\mathcal{F}_{\mathrm{eq}}$ and $\mathcal{A}_{\mathrm{eq}}^{\nu}$, 
\bel{eq:ke}
k^{\mu} \partial_{\mu} \mathcal{F}_{\mathrm{eq}}(x, k)=0, \quad k^{\mu} \partial_{\mu} \mathcal{A}_{\mathrm{eq}}^{\nu}(x, k)=0, \quad k_{\nu} \mathcal{A}_{\mathrm{eq}}^{\nu}(x, k)=0.
\eel
In the case of global equilibrium  \rfm{eq:ke} are satisfied exactly which in-turn yields the conditions that $\beta_{\mu}$ is a Killing vector, whereas, $\xi$ and $\omega_{\mu \nu}$ are constant, but $\omega_{\mu \nu}$ does not necessarily be equal to thermal vorticity $\varpi_{\mu \nu} = -\frac{1}{2} \left(\p_\mu \beta_\nu - \p_\nu \beta_\mu \right) = \hbox{const}$.
But in the case of local equilibrium \rfm{eq:ke} are not exactly followed, here we follow\cite{Denicol:2012cn} and by permitting $\beta$, $\xi$ and $\omega$ dependence on $x$, we need to have only certain moments in momentum space of the kinetic equations \rfn{eq:ke} which are satisfied, which lead to conservation laws for charge, energy-linear momentum and spin\cite{Florkowski:2018ahw}
\begin{eqnarray} 
\quad\partial_\mu N^\mu = 0, \label{Ncons} \\
\partial_\mu T^{\mu\nu}_{\rm GLW} = 0, \label{Tcons}\\
\partial_\lambda  S_{\rm GLW}^{\lambda, \alpha \beta} =0,
\label{Scons}
\end{eqnarray}
here the baryon current, the energy-momentum and the spin tensors are based on the forms by the de Groot - van Leeuwen - van Weert (GLW)~\cite{DeGroot:1980dk}
\bea
N^\alpha &=& n U^\alpha,\\
 T^{\a\b}_{\rm GLW} &=& (\varepsilon + P ) U^\a U^\b - P g^{\a\b},\\
 S^{\alpha , \beta \gamma }_{\rm GLW}
&=& \cosh(\xi) \LS n_{(0)} U^\alpha \omega^{\beta\gamma}  +  {\cal A}_{(0)} \, U^\a U^\d U^{[\b} \omega^{\gamma]}_{\HP\d}  \right. \\
&& +\left. \, {\cal B}_{(0)} \, \Big( 
U^{[\b} \Delta^{\a\d} \omega^{\gamma]}_{\HP\d}
+ U^\a \Delta^{\d[\b} \omega^{\gamma]}_{\HP\d}
+ U^\d \Delta^{\a[\b} \omega^{\gamma]}_{\HP\d}\Big) \RS,
\eel
where $\Delta^{\a\b} =g^{\a\b} - U^\a U^\b$ is the spatial projection operator which is orthogonal to the hydrodynamic flow 4-vector $U$.\\
For the case of polarization tensor in the leading order,  the baryon number density, the energy density and, the pressure are expressed respectively as
\bea 
n &=& \sinh(\xi)\, \nU(T), \label{n0small} \\
\varepsilon &=& \cosh(\xi) \, \eU(T),\label{e0small}\\ 
P &=& \cosh(\xi) \, \PU(T), \label{P0small} 
\eea
where for spin-less and neutral massive Boltzmann particles, thermodynamical properties are defined by\cite{Florkowski:2010zz}
\beq
\nU(T) &=&  \f{2~T^3}{\pi^2}\,  \hat{m}^2 K_2\left( \hat{m}\right), \label{n0c}\\
\eU(T) &=& \f{2~T^4 }{\pi^2}  \, \hat{m} ^2
 \Big[ 3 K_{2}\left( \hat{m} \right) + \hat{m}  K_{1} \left( \hat{m}  \right) \Big],  \label{e0c}\\
\PU(T) &=& T \, \nU(T) . \label{P0c}
\eeq
 Here, $K_{1} \left( \hat{m}  \right)$ and $K_{2} \left( \hat{m}  \right)$ are modified Bessel functions of 1st and 2nd kind respectively. The thermodynamical quantities ${\cal B}_{(0)} $ and ${\cal A}_{(0)}$ are expressed as
\beq
{\cal B}_{(0)} =-\frac{2}{\hat{m}^2} s_{(0)}(T) ,\qquad
{\cal A}_{(0)}  = -3{\cal B}_{(0)} +2 n_{(0)}(T) 
\eeq
with entropy density $\sU =   \LR\eU+\PU\RR / T$ and $\hat{m}=m/T$. 
%
%
\section{Bjorken expansion set-up}
Since the spin polarization tensor $\omega_{\mu\nu}$ is a 2nd rank asymmetric tensor, so in analogy to the Faraday electromagnetic field strength tensor, it can be written into electric-like ($\kappa$) and magnetic-like ($\omega$) components
\beq
\omega_{\mu\nu} &=& \kappa_\mu U_\nu - \kappa_\nu U_\mu + \epsilon_{\mu\nu\a\b} U^\a \omega^{\b},\label{spinpol1}
\eeq
where $\kappa$ and $\omega$ are 4-vectors, orthogonal to fluid flow vector $U_\mu$. 
%
For longitudinal boost-invariant and transversely homogeneous systems\cite{Bjorken:1982qr,Rezzolla:2013zz}, one can write the following basis vectors 
\beq
U^\a &=& \frac{1}{\tau}\LR t,0,0,z \RR = \LR \cosh(\eta), 0,0, \sinh(\eta) \RR, \nn \\
X^\a &=& \LR 0, 1,0, 0 \RR,\nn\\
Y^\a &=& \LR 0, 0,1, 0 \RR, \nn\\
Z^\a &=& \frac{1}{\tau}\LR z,0,0,t \RR = \LR \sinh(\eta), 0,0, \cosh(\eta) \RR, 
\label{BIbasis}
\eeq
where longitudinal proper time is defined as $\tau = \sqrt{t^2-z^2}$ and, the space-time rapidity is defined as $\eta = \half \ln((t+z)/(t-z))$.
The normalization conditions satisfied by the basis vectors \rfn{BIbasis} are
\begin{eqnarray}
 &&U \cdot U = 1,\nn\\
X \cdot X &=&   Y \cdot Y \,\,=\,\, Z \cdot Z \,\,=\,\, -1,\nn \\ \label{orthXYZ}
X \cdot U  \,\,&=& Y \cdot U\,\, \,\,=\,\, Z \cdot U \,\,=\,\, 0,   \\
X \cdot Y  &=&  Y \cdot Z \,\,=\,\, Z \cdot X \,\,=\,\, 0.  \nn
\end{eqnarray}
Using the fact that  $\kappa$ and $\omega$ are orthogonal to $U_\mu$ and \rfm{orthXYZ}, $\kappa^{\mu}$ and $\omega^{\mu}$ can be written as
\beq
\kappa^\a &=&  C_{\kappa X}(\tau) X^\a + C_{\kappa Y}(\tau) Y^\a + C_{\kappa Z}(\tau) Z^\a,  \nn\\
\omega^\a &=&  C_{\omega X}(\tau) X^\a + C_{\omega Y}(\tau) Y^\a + C_{\omega Z}(\tau) Z^\a, \label{decom}
\eeq
where one can notice that the scalar functions depend only on proper time $(\tau)$.\\
Putting \rfm{decom} in \rf{Scons} and then using the projection method, we project the resulting tensor on different combination of basis vectors $U_\a X_\b$, $U_\a Y_\b$, $U_\a Z_\b$, $Y_\a Z_\b$, $X_\a Z_\b$ and $X_\a Y_\b$, we get the six equations of motions as
\begin{equation}
{\rm diag}\LR
\cal{L}, \cal{L}, \cal{L}, \cal{P}, \cal{P}, \cal{P}\RR \,\,
\Dot{\Cv} ={\rm diag}\LR
{\cal{Q}}_1, {\cal{Q}}_1, {\cal{Q}}_2, {\cal{R}}_1, {\cal{R}}_1, {\cal{R}}_2 \RR\,\,
\Cv, \label{cs}
\end{equation} 
where $\Cv = \LR C_{\kappa X}, C_{\kappa Y}, C_{\kappa Z},  C_{\omega X}, C_{\omega Y}, C_{\omega Z} \RR$, $\dot{(\dots)} \equiv U \cdot \p = \p_\tau$ and  
\beq
{\cal L}(\tau)&=&{\cal A}_1-\frac{1}{2}{\cal A}_2-{\cal A}_3,\nn\\
{\cal P}(\tau)&=&{\cal A}_1,\nn\\
{\cal{Q}}_1(\tau)&=&-\left[\dot{{\cal L}}+\frac{1}{\tau}\left( {\cal L}+ \frac{1}{2}{\cal A}_3\right)\right],\nn\\
 {\cal{Q}}_2(\tau)&=&-\left(\dot{{\cal L}}+\frac{{\cal L}}{\tau}   \right),\nn\\
  {\cal{R}}_1(\tau)&=&-\left[\Dot{\cal P}+\frac{1}{\tau}\left({\cal P} -\frac{1}{2} {\cal A}_3 \right) \right],\nn\\
 {\cal{R}}_2(\tau)&=&-\left(\Dot{{\cal P}} +\frac{{\cal P}}{\tau}\right).\nn
 \label{LPQR}
 \eeq
 with 
 \beq
{\cal A}_1 &=& \cosh(\xi) \LR \nU -  {\cal B}_{(0)} \RR \label{A1} ,\nn\\ 
{\cal A}_2 &=& \cosh(\xi) \LR {\cal A}_{(0)} - 3{\cal B}_{(0)} \RR \nn \label{A2} , \\ 
{\cal A}_3 &=& \cosh(\xi)\, {\cal B}_{(0)}\label{A3},\nn
\eeq
\rfm{cs} implies that the ${C}$ functions evolve independently of each other for the case of Bjorken flow and, ${C}_{\kappa X}$ and ${C}_{\kappa Y}$ (similarly ${C}_{\omega X}$ and ${C}_{\omega Y}$) follows the same form of evolution equations due to the rotational invariance.\\
Charge current conservation \rfn{Ncons} for Bjorken type flow is expressed as
\beq
\frac{dn}{d\tau}+\frac{n}{\tau}=0\label{ns}
\eeq
whereas the energy and linear momentum conservation law \rfn{Tcons} (after projecting on $U$) yields
\beq
\frac{d\varepsilon}{d\tau}+\frac{(\varepsilon+P)}{\tau}=0.\label{ts}
\eeq 
%
\section{Particle spin polarization at freeze-out}
To calculate the mean spin polarization per particle, the following formula is used~\cite{Florkowski:2018ahw}
\beq
\langle\pi_{\mu}\rangle=E_p\frac{d\Pi _{\mu }(p)}{d^3 p}/E_p\frac{d{\cal{N}}(p)}{d^3 p}, \label{pi}
\eeq
with $E_p\frac{d\Pi _{\mu }(p)}{d^3 p}$ being the total value of the Pauli-Luba\'nski vector (after integrating over the freeze-out hypersuface, $\Delta \Sigma_{\lambda }$),
\beq
E_p\frac{d\Pi _{\mu }(p)}{d^3 p} = -\f{ \cosh(\xi)}{(2 \pi )^3 m}
\int
\Delta \Sigma _{\lambda } p^{\lambda } \,
e^{-\beta \cdot p} \,
\tilde{\omega }_{\mu \beta }p^{\beta },  \nn
\eeq
and 
\beq
E_p\frac{d{\cal{N}}(p)}{d^3 p}&=&
\f{4 \cosh(\xi)}{(2 \pi )^3}
\int
\Delta \Sigma _{\lambda } p^{\lambda } 
\,
e^{-\beta \cdot p}, \nn
\eeq
is the total momentum density of both particles and antiparticles with four-momentum given as $p^\lambda = \left( m_T\ch y_p,p_x,p_y,m_T\sh y_p \right)$.\\
After performing the the canonical boost \cite{Leader:2001gr} of \rfn{pi}, we obtain the polarization vector $\langle\pi^{\star}_{\mu}\rangle$ in the local rest frame of the particle as
\beq
\langle\pi^{\star}_{\mu}\rangle=-\frac{1}{8m }\left[\begin{array}{c}
0 \\ \\
\left(\frac{p_x\sh y_p }{b}\right) a_i +
  \LR\frac{ \chi \,p_x \ch y_p  }{b} \RR a_j\!+\!2 C_{\kappa Z} p_y  \!-\!\chi C_{\omega X}{m}_T \\ \\
\left(\frac{p_y\sh y_p }{b}\right)a_i+   \LR\frac{ \chi \,p_y \ch y_p  }{b} \RR a_j\!-\!2 C_{\kappa Z} p_x \!-\!\chi C_{\omega Y}{m}_T \\ \\
 -\left(\frac{m\ch y_p+m_T}{b}\right)a_i
-\LR\frac{\chi \,m\,\sh y_p}{b}\RR a_j \nn\\
\end{array}
\right],\\
\label{PLVPPLPRF}  
\eeq 
with $a_i=\chi\left(C_{\kappa X} p_y-C_{\kappa Y} p_x\right)+2 C_{\omega Z} m_T$, $a_j=C_{\omega X} p_x+C_{\omega Y} p_y$, $b=m_T \ch y_p+m$, and $\CHI=\left( K_{0}\left( \hat{m}_T \right)+K_{2}\left( \hat{m}_T \right)\right)/K_{1}\left( \hat{m}_T \right)$ and  $\hat{m}_T=m_T/T$.
\begin{figure}
\centering
 \includegraphics[width=0.6\textwidth]{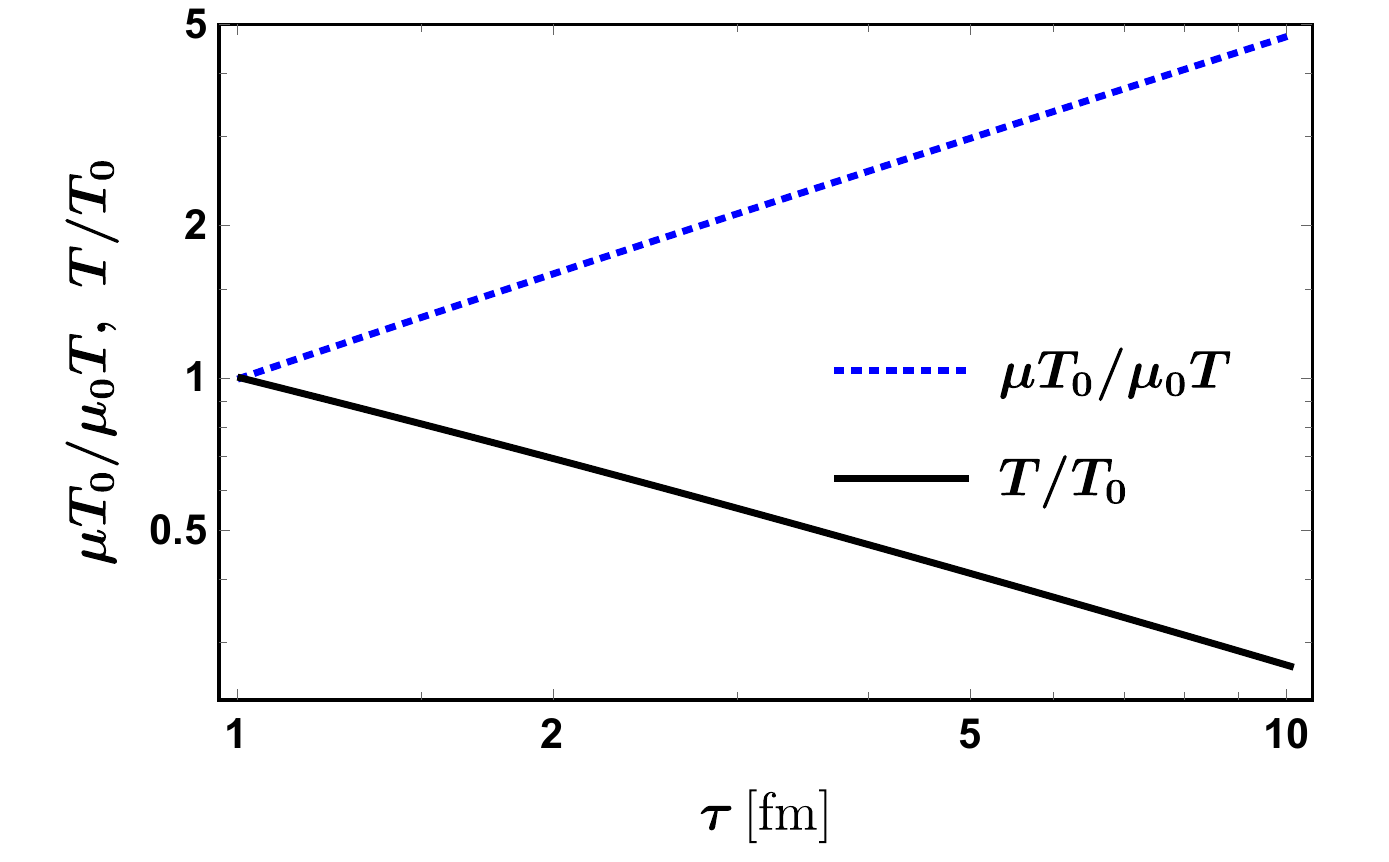} 
\caption{Dependence of the temperature re-scaled by its initial value (solid black line) and the ratio of baryon chemical potential over temperature re-scaled by the initial ratio (dotted blue line) on the proper-time.}
\label{fig:Tmu}
\end{figure}

\begin{figure}
\centering
 \includegraphics[width=0.6\textwidth]{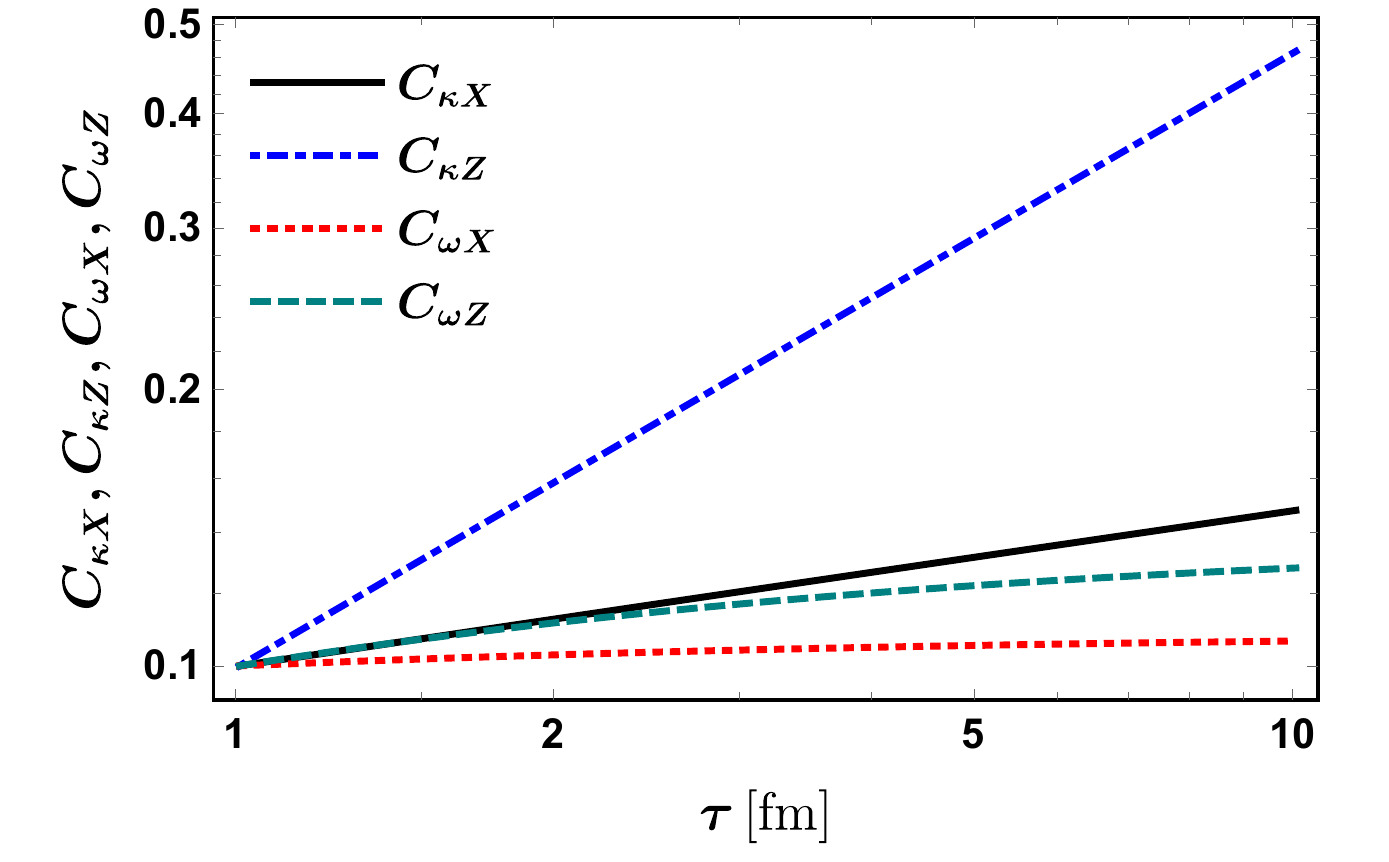} 
\caption{Dependence of scalar functions $C_{\kappa X}$ (solid black line), $C_{\kappa Z}$ (dashed-dotted blue line), $C_{\omega X}$ (dotted red line) and $C_{\omega Z}$ (dashed green line) on the proper-time.}
\label{fig:C}
\end{figure}
%
\section{Results}
\begin{figure}
\centering
 \includegraphics[width=0.32\textwidth]{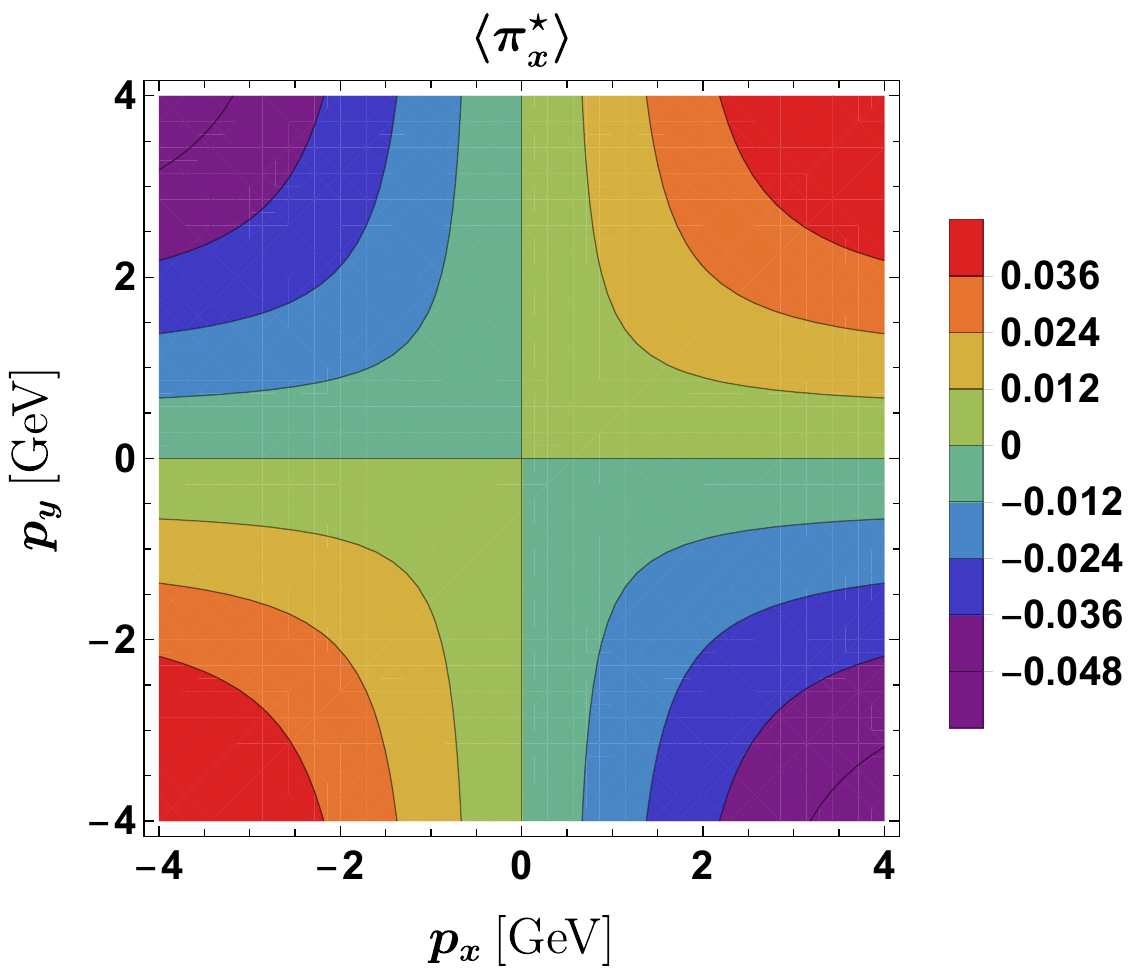}
\includegraphics[width=0.32\textwidth]{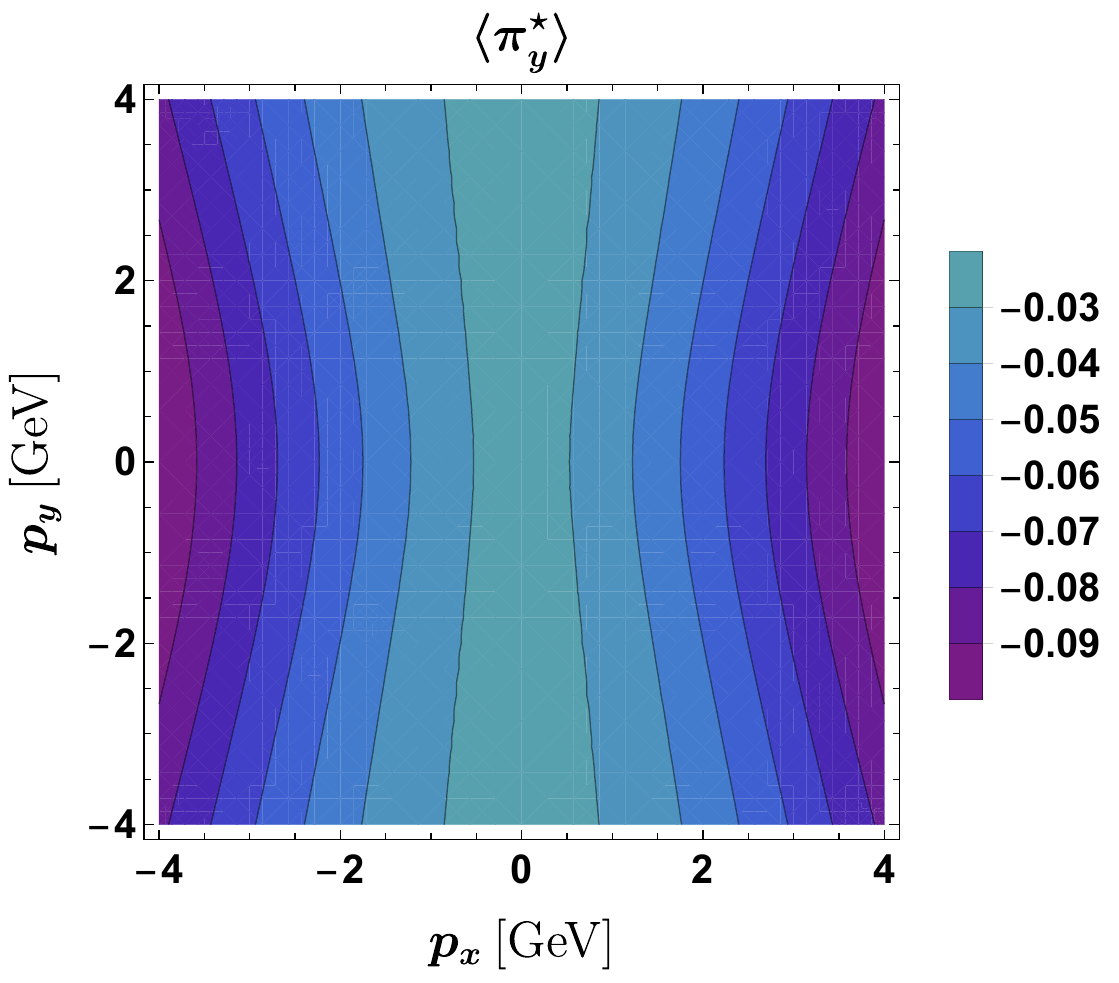}
 \includegraphics[width=0.32\textwidth]{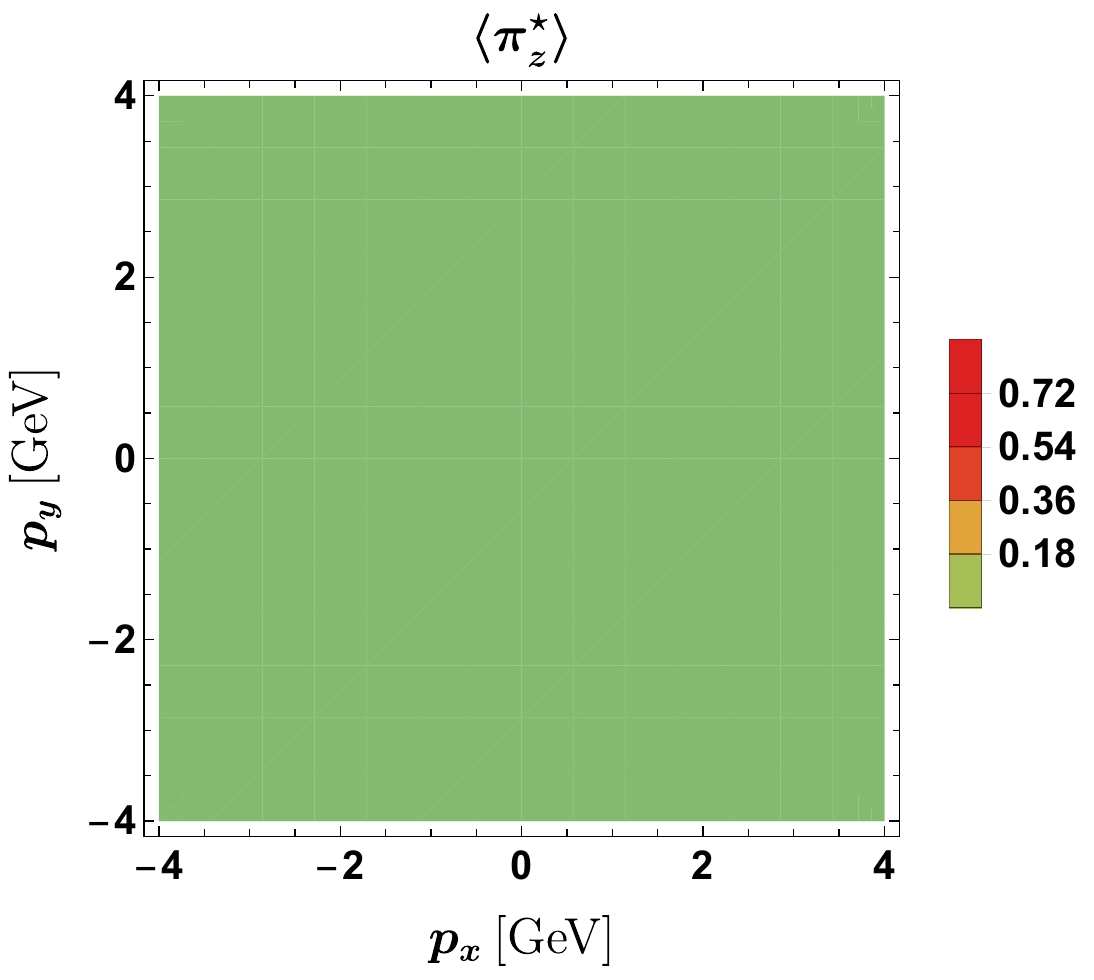}
\caption{Different components of the mean polarization of $\Lambda$ particles in the rest frame of the particle obtained with the initial values $\mu_0=800$~MeV,
$T_0=155$~MeV, $C_{\kappa, 0}=(0,0,0)$ and $ C_{\omega, 0}=(0,0.1,0)$ for $y_p=0$.}
\label{fig:polarization1}
\end{figure}

%
%
Here we show the solutions of the differential equations \rfn{cs}, \rfn{ns}, and \rfn{ts}. System is initialized at the initial proper time $\tau_0 = 1$ fm with initial temperature and the initial baryon chemical potential as $T_0=T(\tau_0)=150$ MeV and $\mu_0=\mu(\tau_0)=800$ MeV, respectively. Here the system is assumed to be formed with $\Lambda$ particles having mass $m=1116$ MeV. In Fig.~\ref{fig:Tmu}, proper-time dependence of temperature and baryon chemical potential is depicted, where the temperature decreases with proper-time, whereas the ratio of baryon chemical potential and temperature increases with proper-time. From Fig.~\ref{fig:C}, proper time dependence of the $C$ functions can be known describing the spin polarization evolution of the system.\\
Using the information of the thermodynamic parameters and $C$ coefficients evolution, we can calculate the different components of the mean polarization vector in the rest frame of the particles $\langle\pi^{\star}_{\mu}\rangle$ at freeze-out, see Fig.~\ref{fig:polarization1}.
We note that $\langle\pi^{\star}_{y}\rangle$ is negative reflecting the system's initial spin polarization. Because of the Bjorken symmetry which we have assumed in our calculations in this article, the longitudinal component ($\langle\pi^{\star}_{z}\rangle$) of the mean polarization vector is vanishing which is not in agreement with the quadrupole structure of the longitudinal component of the spin polarization seen in the experiment. But we note here that $\langle\pi^{\star}_{x}\rangle$ shows quadrupole structure. We however see and note that the Bjorken set-up is very simple to address the measurements done by the experiment.

%
%
\section{Summary}
%
%
We briefly presented the key ingredients of relativistic perfect-fluid hydrodynamics with spin framework initiated recently. From the definitions of kinetic theory for the equilibrium phase space distribution functions in the local equilibrium we obtained the equations of motions for the expansion of the the system. For the case of Bjorken type of flow we investigated the system's spin polarization dynamics, which in turn show that the scalar functions describing the dynamics of the spin polarization evolve independently of each other. These results are used to obtain the particle spin polarization at the freeze-out hypersurface. We however note that within the current simple set-up of Bjorken symmetry experimental measurements cannot be addressed properly.
%
\section*{Acknowledgments}
%
I thank Wojciech Florkowski, Radoslaw Ryblewski and Avdhesh Kumar for inspiring discussions and clarifications. This research is supported in part by the Polish National Science Center Grants No.   2016/23/B/ST2/00717 and No. 2018/30/E/ST2/00432.  
%

\end{document}